\documentclass[aps,pra,superscriptaddress]{revtex4-2}

\usepackage{amsmath,amsfonts,amssymb}
\usepackage{xcolor,graphicx}
\usepackage{braket}
\usepackage{float}
\usepackage[pdftex]{hyperref}
\hypersetup{
	colorlinks = true,
	linkcolor = blue,
	anchorcolor = blue,
	citecolor = red,
	filecolor = blue,
	urlcolor = blue}
\usepackage{subfig}
\usepackage{overpic}

\usepackage{listings}
\usepackage{xcolor}
\usepackage[utf8]{inputenc}
\usepackage{float}

\lstdefinestyle{python}{
    language=Python,
    backgroundcolor=\color{white},   
    basicstyle=\footnotesize\ttfamily, 
    breaklines=true,                  
    captionpos=b,                     
    numbers=left,                     
    numberstyle=\tiny\color{gray},   
    keywordstyle=\color{blue},        
    commentstyle=\color{green!80!black}, 
    stringstyle=\color{red},          
}

\begin{document}

\title{A computational and pedagogical framework for projectile motion using Python visualizations}

\author{L. Hernández-Sánchez}
\email[e-mail: ]{leonardi1469@gmail.com}
\affiliation{Instituto Nacional de Astrofísica Óptica y Electrónica, Calle Luis Enrique Erro No. 1\\ Santa María Tonantzintla, Puebla, 72840, Mexico}
\author{I. Ramos-Prieto}
\affiliation{Instituto Nacional de Astrofísica Óptica y Electrónica, Calle Luis Enrique Erro No. 1\\ Santa María Tonantzintla, Puebla, 72840, Mexico}
\author{F. Soto-Eguibar}
\affiliation{Instituto Nacional de Astrofísica Óptica y Electrónica, Calle Luis Enrique Erro No. 1\\ Santa María Tonantzintla, Puebla, 72840, Mexico}
\author{H.M. Moya-Cessa}
\affiliation{Instituto Nacional de Astrofísica Óptica y Electrónica, Calle Luis Enrique Erro No. 1\\ Santa María Tonantzintla, Puebla, 72840, Mexico}

\date{\today}

\begin{abstract}
Projectile motion is one of the most fundamental problems in introductory physics, offering a clear context to connect algebraic reasoning with conceptual understanding. This work presents a computational and pedagogical framework that combines the analytical formulation of projectile motion with interactive visualizations developed in \texttt{Python}. Using reproducible simulations, the dependence of the maximum height and horizontal range on the launch parameters $(v_0,\theta)$ is examined through trajectory plots, parameter-space maps, and iso-curves. These visual representations reveal non-trivial combinations of initial conditions that yield equivalent outcomes, reinforcing physical intuition and providing an accessible open-source tool for teaching and learning classical mechanics.\\
\vspace{0.5cm}
\textbf{Keys}: projectile motion; computational physics; Python; physics education; visualization.
\end{abstract}
\maketitle

\section{Introduction}\label{Introduction}

Projectile motion remains one of the most fundamental and illustrative topics in introductory kinematics, as it enables students to explore the relationships among the variables that govern the bidimensional motion of a body under uniform gravity. Through this model, students analyze how the launch angle and the initial velocity determine the trajectory, maximum height, and horizontal range of a projectile. 
However, this topic is often taught procedurally, emphasizing numerical substitution in formulas rather than fostering conceptual reasoning about the physical meaning of the results or the interdependence between the involved parameters~\cite{AlonsoFinn_1992,Halliday_2014,Serway_2019}.

From a pedagogical point of view, it is essential that students develop a conceptual understanding of motion by posing and investigating questions such as: 
\textit{Can different launch angles yield the same range?} 
\textit{Which combinations of angle and velocity produce the same maximum height?} 
or \textit{How do height and range vary simultaneously when the initial conditions are modified?} 
Such inquiries foster physical reasoning and modeling skills, both of which are core components of scientific training~\cite{Redish_2003}. In this context, computational environments such as \texttt{Python} provide a powerful platform for interactive and visual exploration of these relationships, bridging analytical formulations with simulated experiences that enhance conceptual understanding and promote active learning~\cite{Giordano_2006,Backer_2007,Rojas_2009}.

In recent years, computational modeling has become a key strategy in modern physics education, fostering active learning, problem solving abilities, and scientific reasoning. 
Programming environments such as \texttt{Python} offer a simple yet powerful platform for visualizing motion~\cite{Python_2020}, verifying theoretical predictions, and exploring parameter dependence in real time~\cite{Langtangen_2014,Caballero_2012,Caballero_2014}. Moreover, projectile motion provides a natural context to connect analytical formulations with real-world applications, ranging from ballistics and sports to engineering and computer graphics, while encouraging reproducible and exploratory learning through simulation~\cite{Brewe_2008,Paudel_2021}.

The main goal of this work is to present a visual and reproducible approach to teaching projectile motion, integrating analytical derivations and computational simulations in \texttt{Python} to enhance conceptual understanding in undergraduate mechanics. From the classical equations of motion, the analytical expressions for maximum height and horizontal range are numerically implemented to generate trajectories, parameter-space maps, and isocurves that reveal combinations of launch parameters, yielding equivalent physical outcomes. This integration of theory and visualization aims to strengthen conceptual reasoning while providing a reproducible educational tool for physics instruction.

The structure of this article is as follows. Section~\ref{Model} introduces the theoretical framework of projectile motion and derives analytical expressions for the maximum height and horizontal range as functions of the launch parameters. Section~\ref{Visualization} describes the computational implementation developed in \texttt{Python}, presenting the simulated trajectories obtained for different initial velocities and launch angles. Section~\ref{Parametric} extends the analysis to the parameter space, constructing maps and isocurves in the $(\theta,v_0)$ plane that reveal combinations of parameters leading to equivalent physical outcomes. Section~\ref{Discussion} discusses the didactic implications of these results, highlighting their pedagogical relevance and the role of computational modeling in physics education. Finally, Section~\ref{Conclusions} summarizes the main findings and outlines possible extensions for classroom and laboratory applications.

\section{Theoretical model and fundamentals of projectile motion}\label{Model}

\begin{figure}[H]
\centering
\includegraphics[width=0.55\linewidth]{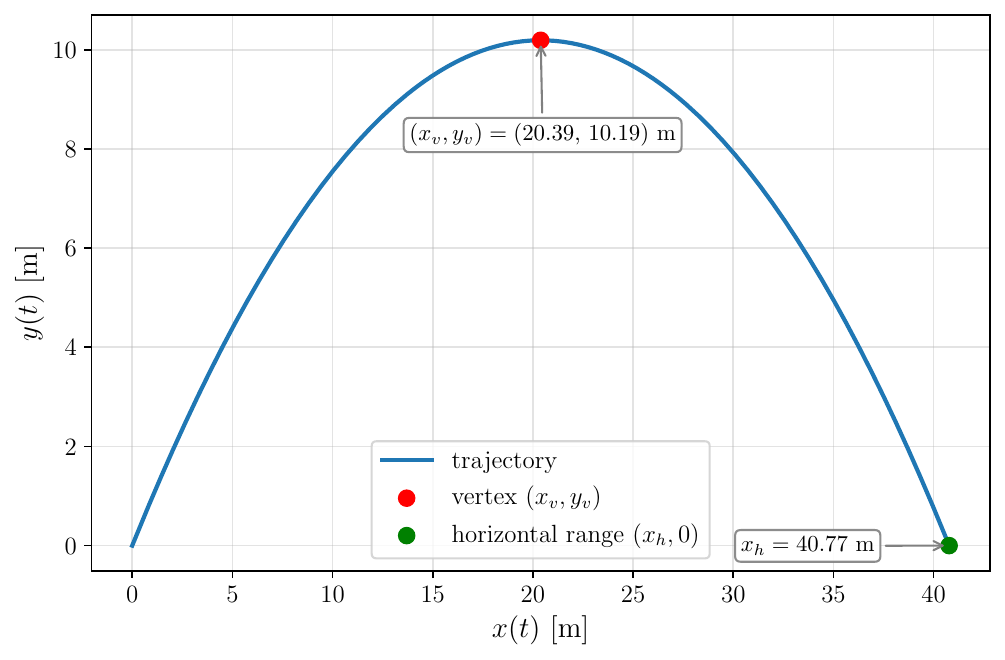}
\caption{Projectile motion for $v_0=20\,\mathrm{m/s}$ and $\theta=45^\circ$. 
The plot shows $x(t)$ and $y(t)$ as functions of time, illustrating the parabolic trajectory under uniform gravity.}
\label{fig:trayectoria}
\end{figure}

Projectile motion describes the trajectory of an object launched with an initial velocity $v_0$ at an angle $\theta$ with respect to the horizontal. 
The velocity vector can be decomposed into two components: a horizontal component which remains constant during the motion and a vertical component, which is affected by the acceleration due to gravity. 
As a result of these two simultaneous motions, the object follows a parabolic path, as illustrated in Figure~\ref{fig:trayectoria}.

These two motions, the horizontal along the $x$-axis and the vertical along the $y$-axis, are independent but connected through time $t$. 
Assuming that the air resistance is negligible, the kinematic equations that describe the position of the projectile at any time $t$ are the following
\begin{subequations}
\begin{align}
x(t) &= v_0 \cos(\theta)\,t, \label{eq:1a}\\
y(t) &= v_0 \sin(\theta)\,t - \frac{1}{2}g\,t^2, \label{eq:1b}
\end{align}
\end{subequations}
where $x(t)$ and $y(t)$ denote horizontal and vertical displacements, respectively; $v_0$ is the initial speed, $\theta$ is the launch angle, and $g = 9.81\,\mathrm{m/s^2}$ is the gravitational acceleration near the surface of the Earth.

\medskip
To express the trajectory in the $xy$ plane, time can be eliminated from Eqs.~\eqref{eq:1a}–\eqref{eq:1b}. 
From Eq.~\eqref{eq:1a}, one obtains
\begin{equation}\label{tiempo}
    t = \frac{x}{v_0 \cos(\theta)}.
\end{equation}
Substituting this result into Eq.~\eqref{eq:1b} gives the vertical position as a function of the horizontal position:
\begin{equation}\label{y_x}
    y(x) = \tan(\theta)\,x - \frac{g}{2v_{0}^{2}\cos^{2}(\theta)}\,x^{2}.
\end{equation}

Equation~\eqref{y_x} has the general quadratic form
\begin{equation}\label{eq_cuadratica}
    y(x) = A x^{2} + Bx + C,
\end{equation}
which characterizes a parabolic trajectory, where the coefficients are:
\begin{subequations}
\begin{align}
A &= -\frac{g}{2v_{0}^{2}\cos^{2}(\theta)}, \label{A}\\
B &= \tan(\theta), \label{B}\\
C &= 0. \label{C}
\end{align}
\end{subequations}

\medskip
From Eq.~\eqref{eq_cuadratica}, the parabola vertex, corresponding to the maximum height of the projectile, can be found by setting the derivative $y'(x)$ equal to zero:
\begin{equation}\label{eq:dy}
y'(x) = 2A x + B = 0.
\end{equation}
Thus, the horizontal coordinate of the vertex is
\begin{equation}\label{eq:xv}
x_{v} = -\frac{B}{2A}.
\end{equation}
Substituting this result into Eq.~\eqref{eq_cuadratica} gives the maximum height:
\begin{equation}\label{eq:yv_general}
y_{v} = A\!\left(-\frac{B}{2A}\right)^{2} + B\!\left(-\frac{B}{2A}\right) = - \frac{B^{2}}{4A}.
\end{equation}

Replacing the coefficients from Eqs.~\eqref{A}–\eqref{C}, one obtains the well-known analytical relations:
\begin{equation}
x_{v} = \frac{v_{0}^{2}\sin(2\theta)}{2g}, \qquad
y_{v} = \frac{v_{0}^{2}\sin^{2}\theta}{2g},
\end{equation}
where $y_v$ represents the maximum height and $x_v$ the horizontal position at which it occurs.

The horizontal range of the projectile is obtained by setting $y(x) = 0$, which yields two roots: $x = 0$ (the launch point) and $x = -\tfrac{B}{A}$ (the landing point). 
The total horizontal displacement, or range, is therefore
\begin{equation}
x_{h} = -\frac{B}{A} = \frac{v_{0}^{2}\sin(2\theta)}{g} = 2x_{v}.
\end{equation}

Finally, Eq.~\eqref{tiempo} expresses the time required for the projectile to reach a given horizontal position $x$. 
From an analytical perspective, the total flight time can be obtained by substituting $x = x_h$ into Eq.~\eqref{tiempo}, yielding
\begin{equation}
T = \frac{2v_0 \sin(\theta)}{g}.
\end{equation}

\medskip
The analytical treatment above establishes the mathematical dependence of the maximum height $y_{v}$ and the horizontal range $x_h$ on the initial conditions. 
Beyond its physical relevance, this formulation has considerable didactic value, since it connects algebraic reasoning with visual interpretation through computational tools. 
Figure~\ref{fig:trayectoria} illustrates these relationships for the case $v_0 = 20\,\mathrm{m/s}$ and $\theta = 45^\circ$, where the analytical parameters $x_v$, $y_v$, and $x_h$ are clearly identified in the numerical plot. 
The complete \texttt{Python} code used to generate this figure is included in Appendix, presented step by step to facilitate its reproduction and to guide students through the process of computing and visualizing the relevant parameters of projectile motion.

Once these relations are understood, a natural pedagogical question arises about the interplay between launch angle and initial velocity: 
can different combinations of these parameters yield equivalent trajectories, the same maximum height, or identical horizontal range? 
Although such relationships are not straightforward to analyze algebraically, computational visualization offers an intuitive means to explore them. 
In the next section, these results are extended by systematically implementing the equations of motion in \texttt{Python}, generating trajectories, parameter maps, and iso-curves that emphasize the conceptual connection between analytical derivations and numerical modeling.

\section{Computational visualization}\label{Visualization}

Based on the analytical relations derived in Section~\ref{Model}, the equations of motion \eqref{eq:1a}–\eqref{eq:1b} and the parabolic form \eqref{y_x} were implemented in \texttt{Python} to visualize the dependence of the projectile trajectory on the launch parameters. 
This computational stage serves as a bridge between the mathematical formulation and its physical interpretation, allowing students to explore how each variable modifies the shape and extent of the motion through direct visualization. 
The complete \texttt{Python} code used for this analysis, including the generation of all trajectories and parameter maps presented in this work, is provided in the Appendix.

\begin{figure}[H]
\centering

\subfloat[Variation with launch angle for fixed $v_0=20\,\mathrm{m/s}$.%
\label{fig:traj_angle}]{
  \includegraphics[width=0.55\linewidth]{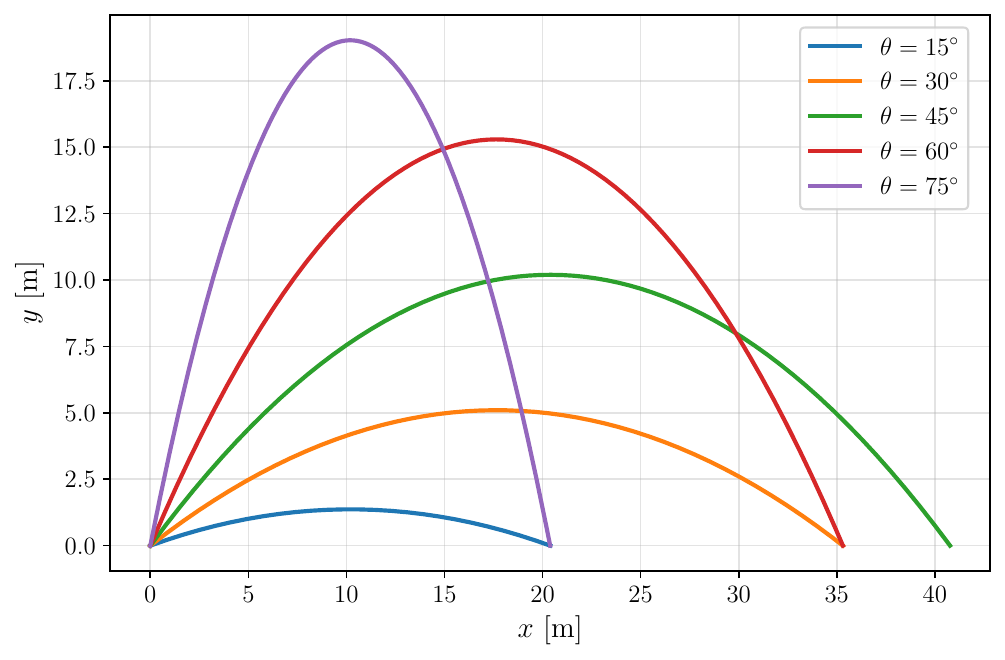}
}\\[0.3cm]

\subfloat[Variation with initial speed for fixed $\theta=45^\circ$.%
\label{fig:traj_speed}]{
  \includegraphics[width=0.55\linewidth]{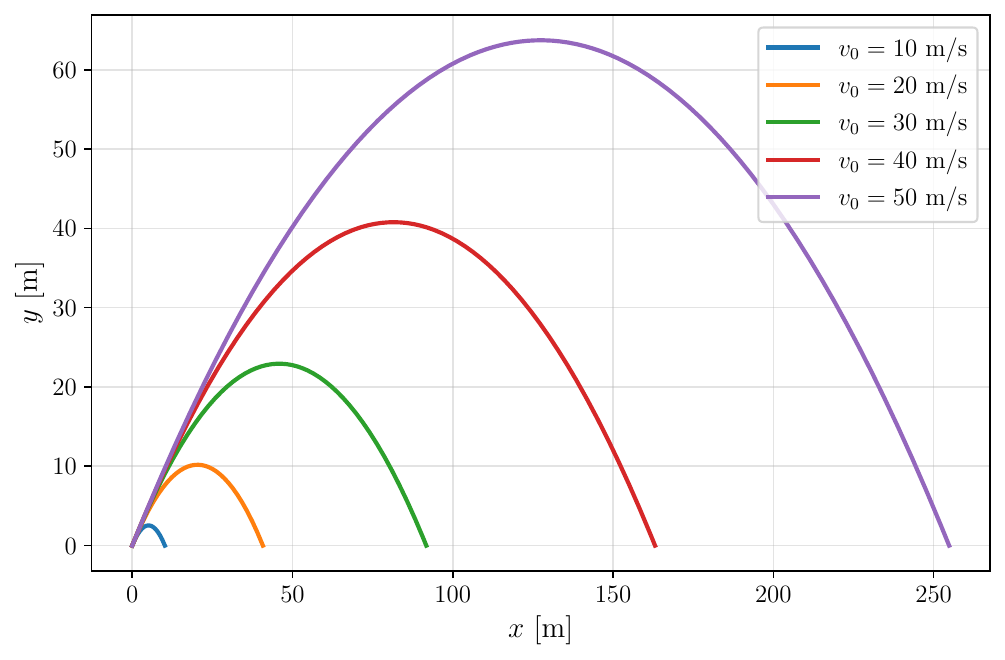}
}
\caption{Projectile trajectories obtained from the analytical equations implemented in \texttt{Python}, illustrating how variations in the launch parameters modify the parabolic path.}
\label{fig:multi_traj}
\end{figure}
\medskip
Figure~\ref{fig:multi_traj} presents two sets of trajectories computed from the analytical expressions using the same numerical procedure described in the appendix. 
In Fig.~\ref{fig:multi_traj}(a), the initial velocity is fixed at $v_0 = 20\,\mathrm{m/s}$ while the launch angle $\theta$ is varied. 
The results clearly reveal the well-known symmetry of the horizontal range described by Eq.~\eqref{y_x}: 
two complementary angles, $\theta$ and $(90^\circ-\theta)$, produce identical horizontal displacements $x_h$ but different maximum heights $y_{v}$. 
This symmetry arises from the trigonometric dependence of the motion, since the vertical component of the velocity, proportional to $\sin(\theta)$, increases up to $45^\circ$ and decreases symmetrically beyond this value. 
Consequently, the same range can be achieved with two different angles, but the height, proportional to $\sin^2(\theta)$, is unique for each launch condition.

\medskip
In contrast, Fig.~\ref{fig:multi_traj}(b) shows the trajectories obtained by varying the initial speed $v_0$ while keeping the launch angle fixed at $\theta = 45^\circ$. 
As predicted by Eqs.~\eqref{y_x} and \eqref{eq:xv}–\eqref{eq:yv_general}, both the horizontal range and the maximum height increase quadratically with $v_0$, resulting in parabolic curves of the same shape but on different scales. 
Unlike the case of angular variation, no two distinct initial speeds can produce the same maximum height or horizontal range, since both quantities depend explicitly on $v_0^2$. 
This behavior highlights the deterministic character of projectile motion and emphasizes the distinct physical roles of angle and speed in shaping the trajectory.

\medskip
These computational visualizations not only confirm the analytical relationships derived earlier but also provide an effective pedagogical tool to explore the dependence of parameters. 
By adjusting the numerical inputs in the provided \texttt{Python} script, students can reproduce the plots, verify the symmetry of the range, and quantify how the trajectory scales with the initial velocity. 
This hands-on experiment promotes active learning, strengthens the connection between theory and computation, and encourages the development of quantitative reasoning and modeling skills within undergraduate mechanics.

\section{Parameter-space analysis}\label{Parametric}

Based on the trajectory analysis presented in Section~\ref{Visualization}, the present section explores the global behavior of projectile motion in the parameter space defined by the launch angle $\theta$ and the initial velocity $v_0$. 
The analytical relations derived in Section~\ref{Model} for the horizontal range and maximum height,
\begin{equation}
x_h = \frac{v_0^2 \sin(2\theta)}{g}, 
\qquad
y_{v} = \frac{v_0^2 \sin^2(\theta)}{2g},
\label{eq:xy_global}
\end{equation}
provide the mathematical foundation for this analysis. 
Both quantities depend quadratically on $v_0$, but their angular dependence differs:
$x_h$ reaches its maximum at $\theta = 45^\circ$, while $y_{v}$ increases monotonically with $\theta$, attaining its maximum at $\theta = 90^\circ$. 
These expressions make it possible to visualize how changes in the launch parameters affect the main physical observables of the projectile.

\begin{figure}[H]
\centering

\subfloat[Horizontal range $x_h(\theta,v_0)$ with iso-line for the reference case.%
\label{fig:param_range}]{
  \includegraphics[width=0.55\linewidth]{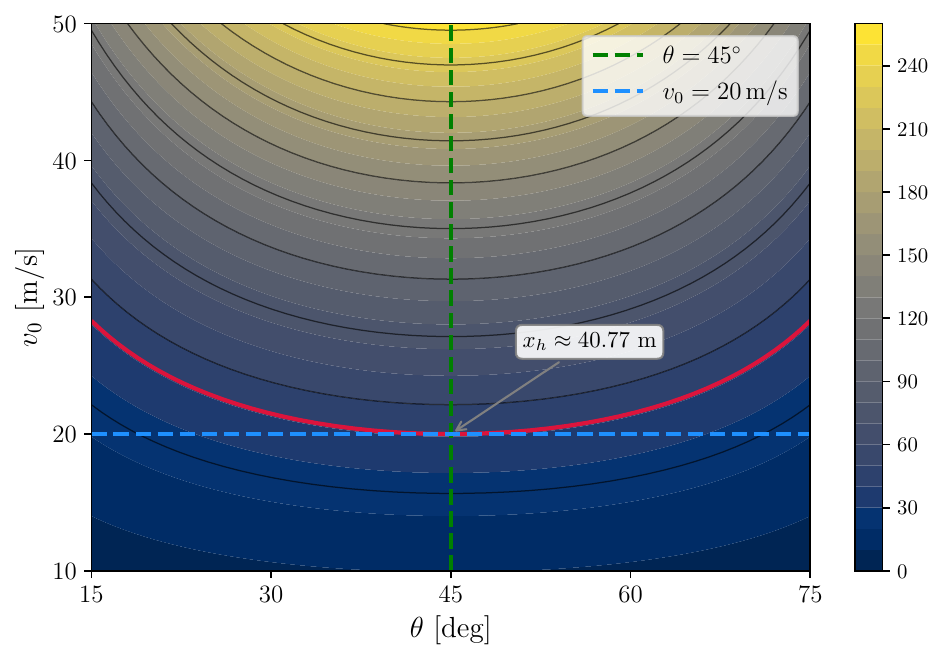}
}\\[0.3cm]

\subfloat[Maximum height $y_{v}(\theta,v_0)$ with iso-line for the same reference case.%
\label{fig:param_height}]{
  \includegraphics[width=0.55\linewidth]{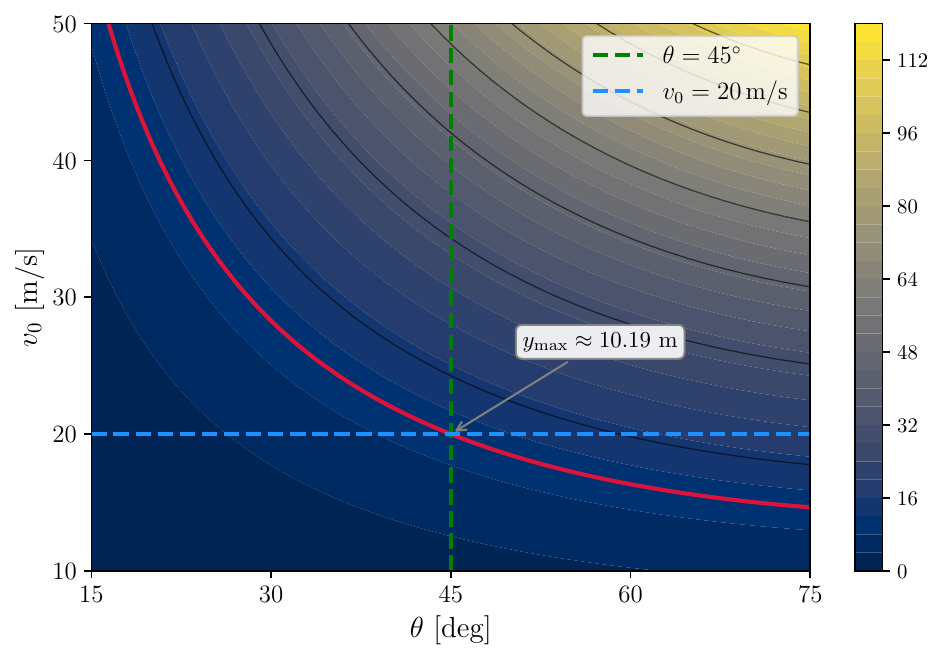}
}\\[0.3cm]

\subfloat[Iso-curves $R=R^*$ and $H=H^*$ in the $(\theta,v_0)$ plane, with the intersection marking $(45^\circ,20\,\mathrm{m/s})$.%
\label{fig:param_iso}]{
  \includegraphics[width=0.55\linewidth]{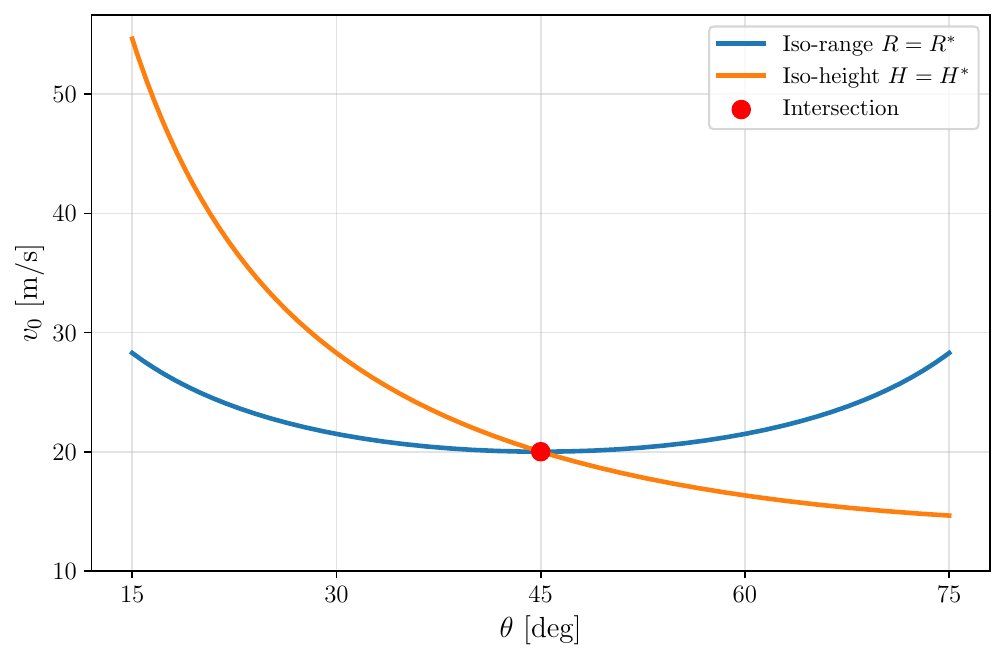}
}

\caption{Parameter-space analysis of projectile motion. 
(a)~Horizontal range and (b)~maximum height as functions of $(\theta,v_0)$; in both panels, the red curve highlights the iso-level corresponding to the reference configuration ($v_0=20\,\mathrm{m/s}$, $\theta=45^\circ$). 
(c)~The corresponding iso-curves $R=R^*$ and $H=H^*$ shown together, emphasizing their intersection at the reference point.}
\label{fig:param_maps}
\end{figure}

\medskip
Figures~\ref{fig:param_maps}(a)–(c) summarize this parameter-space exploration. 
Panels~(a) and~(b) show color maps of $x_h(\theta,v_0)$ and $y_{v}(\theta,v_0)$ calculated from Eq.~\eqref{eq:xy_global} using the same numerical procedure documented in the Appendix. 
In both maps, the color scale represents the corresponding magnitude and the contour lines represent regions of equal values. 
The red curve in each case highlights the \textbf{ isoline} associated with the reference configuration analyzed earlier ($v_0 = 20\,\mathrm{m/s}$, $\theta = 45^\circ$), marking all pairs $(\theta,v_0)$ that reproduce the same horizontal range or the same maximum height as in that case.

\medskip
Panel~(a) reveals the degeneracy of the horizontal range with respect to the launch angle: two complementary angles, $\theta$ and $(90^\circ-\theta)$, produce identical $x_h$. 
This symmetry follows from the trigonometric dependence $\sin(2\theta)$ in Eq.~\eqref{eq:xy_global}, which remains invariant under transformation $\theta \to 90^\circ-\theta$. 
By contrast, panel~(b) shows that $y_{v}$ depends on $\sin^2(\theta)$ and therefore grows monotonically from $0$ to $90^\circ$, implying a unique maximum height for each angle (without angular degeneracy). 
These differences visually emphasize how the horizontal and vertical components of the initial velocity contribute independently to the parabolic trajectory.

\medskip
To isolate these level sets, panel~(c) displays only the iso-curves corresponding to the same target values $R^*$ and $H^*$ (those of the reference configuration) in the $(\theta,v_0)$ plane. 
From Eq.~\eqref{eq:xy_global}, the corresponding analytical forms of these isocurves can be written as:
\begin{equation}
v_0 = \sqrt{\frac{g\,x_h}{\sin(2\theta)}}, 
\qquad
v_0 = \sqrt{\frac{2g\,y_{v}}{\sin^2(\theta)}}.
\label{eq:v0_iso}
\end{equation}
The intersection point of the two curves (marked in red) corresponds exactly to $(\theta,v_0)=(45^\circ,20\,\mathrm{m/s})$, thus linking the trajectory-level analysis of Section~\ref{Visualization} with the global parameter space picture. 
The first iso-curve ($R=R^*$) reflects the angular symmetry of the horizontal range, while the second ($H=H^*$) illustrates the monotonic nature of the maximum height.

\medskip
This parametric visualization provides an integrative framework that connects algebraic reasoning, physical interpretation, and computational exploration. 
It enables students to recognize how different combinations of $\theta$ and $v_0$ can produce equivalent or distinct outcomes, reinforcing conceptual understanding through direct numerical experimentation. 
Reproducing these figures using the \texttt{Python} scripts included in the appendix promotes active learning, allowing students to test predictions, explore new parameter regions, and develop an intuitive grasp of the mathematical relationships governing projectile motion.

\section{Discussion and didactic implications}\label{Discussion}

The combination of analytical derivations and computational visualization developed in this work provides a powerful pedagogical framework for teaching classical mechanics. 
By constructing trajectory plots and parameter-space maps directly from the analytical equations, students gain a deeper appreciation of how mathematical expressions translate into observable physical behavior. 
Such representations facilitate the transition from symbolic manipulation to conceptual understanding, reinforcing the relationship between geometry, trigonometry, and motion.

\medskip
From a didactic perspective, these visualizations support inquiry-based learning strategies by allowing students to formulate and test hypotheses about the dependence of projectile motion on its initial conditions. 
For example, learners can modify the values of $\theta$ and $v_0$ in the provided \texttt{Python} scripts and immediately observe how the parabolic trajectory, horizontal range and maximum height change. 
This interactive approach strengthens understanding of physical principles such as symmetry, conservation of energy, and vector decomposition, while simultaneously fostering computational literacy. 
Furthermore, since the codes are based on open source tools, they can be easily integrated into laboratory sessions or virtual environments without the need for specialized software.

\medskip
Beyond their pedagogical utility, the results have direct relevance in a variety of physical and technological contexts. 
Projectile motion is not only a classical problem in kinematics but also a fundamental model in engineering, sports science, and computer graphics. 
For instance, the equations and visualizations presented here can be adapted to simulate the trajectory of a soccer ball under different launch angles, to optimize the range of fluid or particle ejection systems, or to program realistic motion in animation and game design. 
In all these cases, the same mathematical structure applies and the isocurves of constant range or height serve as intuitive design tools for predicting motion outcomes under varying conditions.

\medskip
Thus, the integration of computational modeling into introductory mechanics goes beyond numerical simulation: it provides an experimental-like environment for exploration, encourages active learning, and bridges theoretical concepts with practical applications. 
The framework developed in this work can be extended to include additional physical effects such as air resistance or energy dissipation, further enriching the analysis and connecting idealized models with real-world phenomena.

\section{Conclusions}\label{Conclusions}

A computational and pedagogical framework for studying projectile motion has been presented, combining analytical derivations with interactive visualizations implemented in \texttt{Python}. 
The approach connects the mathematical formulation of motion with its physical interpretation through trajectory plots, parameter-space maps, and isocurves that reveal equivalent combinations of launch parameters.

\medskip
The results demonstrate that while the horizontal range exhibits angular symmetry for complementary angles, the maximum height is uniquely determined by the launch angle, reflecting the distinct functions of the horizontal and vertical components of motion. 
By visualizing these relationships numerically, students can directly link equations, geometry, and physical intuition, achieving a more comprehensive understanding of the problem.

\medskip
From an educational point of view, this methodology exemplifies how open source computational tools can be integrated into physics curricula to promote active learning, reproducible research practices, and conceptual reasoning. 
The use of \texttt{Python} facilitates accessibility and transparency, enabling both instructors and students to reproduce, modify, and extend the results independently.

\medskip
Finally, the framework established here can serve as a foundation for future developments, including the incorporation of dissipative forces, variable gravity, or stochastic perturbations, thus connecting the simplicity of projectile motion with the complexity of real physical systems. 
Through such extensions, computational modeling continues to emerge as a central tool in the teaching, learning, and application of modern physics.

\appendix
\section*{Appendix: Python scripts for projectile motion simulations}\label{AppendixA}

This appendix compiles all the \texttt{Python} scripts used in this work. Each script corresponds to a specific figure and is presented step by step, with comments to facilitate its understanding and reproduction in classroom or computational laboratory environments.

\bigskip
\noindent\textbf{Figure~1. Projectile motion for $v_0 = 20$ m/s and $\theta = 45^\circ$}

\begin{lstlisting}[style=python,caption={Code for Figure~1 (single-trajectory simulation).},label={lst:fig1}]
import numpy as np
import matplotlib as mpl
import matplotlib.pyplot as plt

# LaTeX text rendering (comment out if LaTeX is not installed)
mpl.rcParams['text.usetex'] = True
mpl.rcParams['font.family'] = 'serif'

# Physical constants and parameters
g = 9.81                 # m/s^2
theta_deg = 45           # fixed angle
theta = np.deg2rad(theta_deg)
v0_list = [10, 20, 30, 40, 50]  # initial speeds to compare
Npts = 400

# Coefficient B of y(x) = A x^2 + B x + C, with C = 0
B = np.tan(theta)

# -------- Figure --------
plt.figure(figsize=(6.8, 4.5))  # good for double-column width

for v0 in v0_list:
    A = - g / (2.0 * v0**2 * np.cos(theta)**2)

    # Range (non-trivial root of y=0): x_h = -B/A
    x_h = - B / A

    # x-grid up to landing
    x = np.linspace(0.0, x_h, Npts)

    # Trajectory: y(x) = A x^2 + B x  (C = 0)
    y = A * x**2 + B * x
    y = np.clip(y, 0.0, None)  # avoid tiny negatives from rounding

    # Label using LaTeX (no unicode)
    plt.plot(x, y, linewidth=2, label=rf'$v_0={v0}\ \mathrm{{m/s}}$')

# Axes, ticks, legend
plt.xlabel(r'$x\ \mathrm{[m]}$', fontsize=14)
plt.ylabel(r'$y\ \mathrm{[m]}$', fontsize=14)
plt.xticks(fontsize=12)
plt.yticks(fontsize=12)

# Legend inside, top-right (adjust as needed)
plt.legend(loc='upper right', fontsize=12, frameon=True)
plt.grid(True, alpha=0.35)
plt.tight_layout()
plt.savefig('Trajectories_by_speed.pdf', dpi=400, bbox_inches='tight')  # optional
plt.show()
\end{lstlisting}

\medskip
\noindent\textbf{Figure~2 — Variation of the trajectory with launch angle}

\begin{lstlisting}[style=python,caption={Code for Figure~2 (trajectories for different launch angles).},label={lst:fig2}]

import numpy as np
import matplotlib as mpl
import matplotlib.pyplot as plt

# LaTeX text rendering (comment out if LaTeX is not installed)
mpl.rcParams['text.usetex'] = True
mpl.rcParams['font.family'] = 'serif'

# Physical constants and parameters
g = 9.81          # m/s^2
v0 = 20.0         # m/s
angles_deg = [15, 30, 45, 60, 75]
Npts = 500

# -------- Figure --------
plt.figure(figsize=(6.8, 4.5))  # good for double-column width

for theta_deg in angles_deg:
    theta = np.radians(theta_deg)

    # Quadratic coefficients of y(x) = A x^2 + B x + C, with C = 0
    A = - g / (2.0 * v0**2 * np.cos(theta)**2)
    B = np.tan(theta)
    C = 0.0

    # Range (non-trivial root of y=0): x_h = -B/A
    x_h = - B / A

    # x-grid truncated exactly at the landing point
    x = np.linspace(0.0, x_h, Npts)

    # Trajectory
    y = A * x**2 + B * x + C
    y = np.clip(y, 0.0, None)  # avoid tiny negative values due to rounding

    # Label using LaTeX (no unicode)
    plt.plot(x, y, linewidth=2, label=rf'$\theta={theta_deg:.0f}^\circ$')

# Axes, ticks, legend
plt.xlabel(r'$x\ \mathrm{[m]}$', fontsize=14)
plt.ylabel(r'$y\ \mathrm{[m]}$', fontsize=14)
plt.xticks(fontsize=12)
plt.yticks(fontsize=12)

# Legend outside (top-right), avoids covering curves
plt.legend(loc='upper left', bbox_to_anchor=(0.77, 1.0), fontsize=12, frameon=True)
plt.grid(True, alpha=0.35)
plt.tight_layout()
plt.savefig('Trajectories_by_angle.pdf', dpi=400, bbox_inches='tight')
plt.show()
\end{lstlisting}

\medskip
\noindent\textbf{Figure~3 — Variation of the trajectory with initial velocity}

\begin{lstlisting}[style=python,caption={Code for Figure~3 (trajectories for different initial speeds).},label={lst:fig3}]

import numpy as np
import matplotlib as mpl
import matplotlib.pyplot as plt

# LaTeX text rendering (comment out if LaTeX is not installed)
mpl.rcParams['text.usetex'] = True
mpl.rcParams['font.family'] = 'serif'

# Physical constants and parameters
g = 9.81                 # m/s^2
theta_deg = 45           # fixed angle
theta = np.deg2rad(theta_deg)
v0_list = [10, 20, 30, 40, 50]  # initial speeds to compare
Npts = 400

# Coefficient B of y(x) = A x^2 + B x + C, with C = 0
B = np.tan(theta)

# -------- Figure --------
plt.figure(figsize=(6.8, 4.5))  # good for double-column width

for v0 in v0_list:
    A = - g / (2.0 * v0**2 * np.cos(theta)**2)

    # Range (non-trivial root of y=0): x_h = -B/A
    x_h = - B / A

    # x-grid up to landing
    x = np.linspace(0.0, x_h, Npts)

    # Trajectory: y(x) = A x^2 + B x  (C = 0)
    y = A * x**2 + B * x
    y = np.clip(y, 0.0, None)  # avoid tiny negatives from rounding

    # Label using LaTeX (no unicode)
    plt.plot(x, y, linewidth=2, label=rf'$v_0={v0}\ \mathrm{{m/s}}$')

# Axes, ticks, legend
plt.xlabel(r'$x\ \mathrm{[m]}$', fontsize=14)
plt.ylabel(r'$y\ \mathrm{[m]}$', fontsize=14)
plt.xticks(fontsize=12)
plt.yticks(fontsize=12)

# Legend inside, top-right (adjust as needed)
plt.legend(loc='upper right', fontsize=12, frameon=True)
plt.grid(True, alpha=0.35)
plt.tight_layout()
plt.savefig('Trajectories_by_speed.pdf', dpi=400, bbox_inches='tight')  # optional
plt.show()
\end{lstlisting}

\medskip
\noindent\textbf{Figure~4 — Horizontal range as a function of $v_0$ and $\theta$}

\begin{lstlisting}[style=python,caption={Code for Figure~4 (range map and contours).},label={lst:fig4}]

import numpy as np
import matplotlib as mpl
import matplotlib.pyplot as plt

# LaTeX text rendering (comment out if LaTeX is not installed)
mpl.rcParams['text.usetex'] = True
mpl.rcParams['font.family'] = 'serif'

# Physical constants
g = 9.81  # m/s^2

# Parameter grids (requested ranges)
theta_deg = np.linspace(15.0, 75.0, 500)  # degrees
v0 = np.linspace(10.0, 50.0, 500)         # m/s

# Mesh: X-axis = theta (deg), Y-axis = v0 (m/s)
TH, V0 = np.meshgrid(np.deg2rad(theta_deg), v0)  # TH in radians

# Horizontal range
R = (V0**2 * np.sin(2.0 * TH)) / g

# Reference point to highlight and to build the iso-range curve
theta_mark_deg = 45.0
v0_mark = 20.0
theta_mark = np.deg2rad(theta_mark_deg)
R_mark = (v0_mark**2 * np.sin(2.0 * theta_mark)) / g

# --- Figure ---
plt.figure(figsize=(6.8, 4.5)) # good for double-column width

# Filled contour (theta on X, v0 on Y)
cs = plt.contourf(theta_deg, v0, R, levels=30, cmap='cividis')
cbar = plt.colorbar(cs)

# Iso-contours overlay
plt.contour(theta_deg, v0, R, levels=10, colors='k', linewidths=0.6, alpha=0.6)

# Iso-range curve at R_mark (highlight) ---
plt.contour(theta_deg, v0, R, levels=[R_mark], colors='crimson', linewidths=2.0)

# Guides at theta=45 and v0=20 m/s
plt.axvline(theta_mark_deg, ls='--', color='green', lw=2, label=r'$\theta=45^\circ$')
plt.axhline(v0_mark,        ls='--', color='dodgerblue', lw=2, label=r'$v_0=20\,\mathrm{m/s}$')

# Annotation with the range value at that point
plt.annotate(
    rf'$x_h \approx {R_mark:.2f}\ \mathrm{{m}}$',
    xy=(theta_mark_deg, v0_mark),
    xytext=(theta_mark_deg + 6, v0_mark + 6),
    textcoords='data',
    ha='left', va='bottom', fontsize=11,
    bbox=dict(boxstyle='round,pad=0.3', fc='white', ec='gray', alpha=0.9),
    arrowprops=dict(arrowstyle='->', color='gray', lw=1)
)

# Axes, ticks, legend
plt.xlabel(r'$\theta$ [deg]', fontsize=14)
plt.ylabel(r'$v_0$ [m/s]',   fontsize=14)

# Set ticks: X every 15 deg, Y every 10 m/s
plt.xticks(np.arange(15, 75 + 1, 15), fontsize=12)
plt.yticks(np.arange(10, 50 + 1, 10), fontsize=12)

plt.grid(False)
plt.legend(loc='upper right', fontsize=12, frameon=True)
plt.tight_layout()
plt.savefig('Horizontal_range.pdf', dpi=400, bbox_inches='tight')  # optional
plt.show()
\end{lstlisting}

\medskip
\noindent\textbf{Figure~5 — Maximum height as a function of $v_0$ and $\theta$}

\begin{lstlisting}[style=python,caption={Code for Figure~5 (maximum height map and contours).},label={lst:fig5}]

import numpy as np
import matplotlib as mpl
import matplotlib.pyplot as plt

# LaTeX text rendering (comment out if LaTeX is not installed)
mpl.rcParams['text.usetex'] = True
mpl.rcParams['font.family'] = 'serif'

# Physical constants
g = 9.81  # m/s^2

# Parameter grids (requested ranges)
theta_deg = np.linspace(15.0, 75.0, 500)  # degrees
v0 = np.linspace(10.0, 50.0, 500)         # m/s

# Mesh: X-axis = theta (deg), Y-axis = v0 (m/s)
TH, V0 = np.meshgrid(np.deg2rad(theta_deg), v0)  # TH in radians

# Maximum height: H = v0^2 * sin^2(theta) / (2g)
H = (V0**2 * np.sin(TH)**2) / (2.0 * g)

# Reference point to highlight and to build the iso-height curve
theta_mark_deg = 45.0
v0_mark = 20.0
theta_mark = np.deg2rad(theta_mark_deg)
H_mark = (v0_mark**2 * np.sin(theta_mark)**2) / (2.0 * g)

# --- Figure ---
plt.figure(figsize=(6.8, 4.5)) # good for double-column width

# Filled contour (theta on X, v0 on Y)
cs = plt.contourf(theta_deg, v0, H, levels=30, cmap='cividis')
cbar = plt.colorbar(cs)

# Iso-contours overlay
plt.contour(theta_deg, v0, H, levels=10, colors='k', linewidths=0.6, alpha=0.6)

# Iso-height curve at H_mark (highlight) ---
plt.contour(theta_deg, v0, H, levels=[H_mark], colors='crimson', linewidths=2.0)

# Guides at theta=45 and v0=20 m/s
plt.axvline(theta_mark_deg, ls='--', color='green', lw=2, label=r'$\theta=45^\circ$')
plt.axhline(v0_mark,        ls='--', color='dodgerblue', lw=2, label=r'$v_0=20\,\mathrm{m/s}$')

# Annotation with the max-height value at that point
plt.annotate(
    rf'$y_{{\max}} \approx {H_mark:.2f}\ \mathrm{{m}}$',
    xy=(theta_mark_deg, v0_mark),
    xytext=(theta_mark_deg + 6, v0_mark + 6),
    textcoords='data',
    ha='left', va='bottom', fontsize=11,
    bbox=dict(boxstyle='round,pad=0.3', fc='white', ec='gray', alpha=0.9),
    arrowprops=dict(arrowstyle='->', color='gray', lw=1)
)

# Axes, ticks, legend
plt.xlabel(r'$\theta$ [deg]', fontsize=14)
plt.ylabel(r'$v_0$ [m/s]',   fontsize=14)

# Set ticks: X every 15 deg, Y every 10 m/s
plt.xticks(np.arange(15, 75 + 1, 15), fontsize=12)
plt.yticks(np.arange(10, 50 + 1, 10), fontsize=12)

plt.grid(False)
plt.legend(loc='upper right', fontsize=12, frameon=True)
plt.tight_layout()
plt.savefig('Max_height.pdf', dpi=400, bbox_inches='tight')  # optional
plt.show()
\end{lstlisting}

\medskip
\noindent\textbf{Figure~6 — Iso-curves of constant range and height}

\begin{lstlisting}[style=python,caption={Code for Figure~6 (iso-curves and intersection).},label={lst:fig6}]

import numpy as np
import matplotlib as mpl
import matplotlib.pyplot as plt

# LaTeX text rendering (comment out if LaTeX is not installed)
mpl.rcParams['text.usetex'] = True
mpl.rcParams['font.family'] = 'serif'

# Constants and marked case
g = 9.81
theta_mark_deg = 45.0
v0_mark = 20.0
theta_mark = np.deg2rad(theta_mark_deg)

# Targets R* and H* from the marked case
R_star = (v0_mark**2 * np.sin(2*theta_mark)) / g
H_star = (v0_mark**2 * (np.sin(theta_mark)**2)) / (2*g)

# Angular range (consistent with your maps)
theta_deg = np.linspace(15.0, 75.0, 1200)
theta = np.deg2rad(theta_deg)

# Iso-range
sin2 = np.sin(2*theta)
v0_isoR = np.full_like(theta, np.nan)
mask_R = np.abs(sin2) > 1e-12
v0_isoR[mask_R] = np.sqrt((g * R_star) / sin2[mask_R])

# Iso-height
sin1 = np.sin(theta)
v0_isoH = np.full_like(theta, np.nan)
mask_H = np.abs(sin1) > 1e-12
v0_isoH[mask_H] = np.sqrt(2*g*H_star) / np.abs(sin1[mask_H])

# --- Figure ---
plt.figure(figsize=(6.8, 4.5)) # good for double-column width
plt.plot(theta_deg[mask_R], v0_isoR[mask_R], lw=2.2, label=r'Iso-range $R=R^*$')
plt.plot(theta_deg[mask_H], v0_isoH[mask_H], lw=2.2, label=r'Iso-height $H=H^*$')

# Intersection point (exactly the marked case)
plt.scatter([theta_mark_deg], [v0_mark], s=70, c='red', zorder=3, label='Intersection')

# Axes and formatting
plt.xlabel(r'$\theta$ [deg]', fontsize=14)
plt.ylabel(r'$v_0$ [m/s]',    fontsize=14)
plt.xticks(np.arange(15, 75+1, 15), fontsize=12)
plt.yticks(np.arange(10, 50+1, 10), fontsize=12)

plt.grid(True, alpha=0.35)
plt.legend(loc='best', fontsize=11, frameon=True)
plt.tight_layout()
plt.savefig('Iso_curves.pdf', dpi=400, bbox_inches='tight')
plt.show()
\end{lstlisting}

\bigskip
Each script can be executed directly in environments such as Jupyter Notebook or Google Colab, allowing students to reproduce all the results presented in the article and modify the parameters $v_0$ and $\theta$ to explore new configurations.

\section*{Acknowledgments}
L. Hernández-Sánchez acknowledges financial support from the Secretaría de Ciencia, Humanidades, Tecnología e Innovación (SECIHTI) through a postdoctoral fellowship (CVU No. 736710).

\bibliographystyle{unsrt} 


\end{document}